\documentclass[conference]{IEEEtran}

\usepackage{cite}
\usepackage[pdftex]{graphicx}

\usepackage{balance}
\usepackage{tikz}
\usepackage{amsmath}
\usepackage{multicol}
\usepackage{multirow}
\usepackage{fontawesome}
\usepackage{array}
\usepackage{mathtools}
\usepackage{colortbl}
\usepackage{xspace}
\usepackage{kmath}
\usepackage{booktabs}
\usepackage[hidelinks]{hyperref}
\usepackage{pgfplots}
\pgfplotsset{compat=1.18}
\definecolor{mygray}{HTML}{ededed}
\usepackage{longtable}
\usepackage{subcaption}

\begin{document}

\title{Design of Secure, Privacy-focused, and Accessible E-Payment Applications for Older Adults}

\author{\IEEEauthorblockN{Sanchari Das} \IEEEauthorblockA{\textit{Department of Computer Science, University of Denver, USA}\\ \textit{Information Sciences and Technology Department, George Mason University, USA}\\ \underline{{sanchari.das}@du.edu} \& \underline{{sdas35}@gmu.edu}}}

\IEEEoverridecommandlockouts
\makeatletter\def\@IEEEpubidpullup{6.5\baselineskip}\makeatother
\IEEEpubid{\parbox{\columnwidth}{
    BuildSEC'24 Building a Secure \& Empowered Cyberspace\\
    19-21 December 2024, New Delhi, India\\
    https://www.buildsec.org/\\
}
\hspace{\columnsep}\makebox[\columnwidth]{}}

\maketitle

\begin{abstract}
E-payments are essential for transactional convenience in today’s digital economy and are becoming increasingly important for older adults, emphasizing the need for enhanced security, privacy, and usability. To address this, we conducted a survey-based study with $400$ older adults aged $60$ and above to evaluate a high-fidelity prototype of an e-payment mobile application, which included features such as multi-factor authentication (MFA) and QR code-based recipient addition. Based on our findings, we developed a tailored $\beta$ version of the application to meet the specific needs of this demographic. Notably, approximately 91\% of participants preferred traditional knowledge-based and single-mode authentication compared to expert-recommended MFA. We concluded by providing recommendations aimed at developing inclusive e-payment solutions that address the security, privacy, and usability requirements of older adults.
\end{abstract}


\section{Introduction}
Electronic payment (e-payment) is defined by Dahlberg et al. as ``payments for goods, services, and bills with a mobile device (such as a mobile phone, smartphone, or personal digital assistant (PDA)) by taking advantage of wireless and other communication technologies''~\cite{dahlberg2008past}. According to Statista, the total transaction value in the digital payments market is anticipated to reach US~\textdollar $11.55$ trillion in $2024$~\cite{statista}. This type of payment has greatly permeated today's economic interactions. Moreover, e-payments are made using mobile applications that are utilized by a wide range of users, including those of different ages, with $64\%$ of those in the $58-69$ age group and $80\%$ of those in the $43-58$ age group reporting using e-payment apps~\cite{mckinsey_company_2019}.

The growing acceptance of digital payments highlights their vulnerability to cyberattacks, including replay, man-in-the-middle, impersonation, and unauthorized access techniques~\cite{agarwal2007security,kishnani2023assessing,kishnani2023towards}. Attackers regularly target e-payment applications because of the sensitive data they manage~\cite{preibusch2016shopping,ho_price_2020}. Numerous personally identifiable information (PII) such as names, addresses, and financial information like debit and credit card numbers are frequently accessed by these apps~\cite{filkins2016privacy,dev2018privacy,saglam2022personal,kishnani2023optimizing,noah2022privacy}. By 2025, the global cost of cybercrime is expected to be approximately US\$ $10.5$ trillion, with $200$ zettabytes of data being the primary attack surface~\cite{cybercrimemag_2021}. The usability and design of e-payment apps are just as important as security and privacy issues. Previous research has demonstrated that user perceptions of security and privacy are influenced by the usability of e-payment mobile applications~\cite{ozkan2010facilitating,momenzadeh2021bayesian,gopavaram2021iot,das2019privacy}. However, the user's perspective and usage difficulties are still poorly researched. This is particularly true for understudied populations like older individuals~\cite{das2019don,noah2024aging,joshi2020substituting}, who can particularly benefit from collaborative user-focused work. Thus, to understand further we integrated insights from previous co-design research and a pre-study on security, privacy, usability, and accessibility to develop a high-fidelity prototype, which was then evaluated through a crowdsourced user study with $400$ older adults, leading to a $\beta$ version of the e-payment application.

\section{Method}
\label{sec:ph2_methodology}
\subsection{High-Fidelity Prototype Building}
Based on the input and recommendations from our co-design user study, we built a high-fidelity prototype using Figma~\cite{figma}. This prototype includes several interconnected pages that implement security, privacy, and usability concepts. Usability principles are integrated throughout all the prototype pages, totaling $17$ screens. The prototype prominently features the option to view the privacy policy. Instead of a long policy that users typically do not read, we offer a short, simplified privacy policy screen to assess user engagement. Additionally, unlike most other payment apps, our prototype requires Multi-Factor Authentication (MFA) during login and registration~\cite{jensen2021multi,das2020risk,das2021organizational}. It includes four different MFA options, allowing users to choose their preferred method, thus enhancing security while giving users control over their choices.
\subsubsection{Authentication Flow}
The authentication flow of the prototype begins with a standard login screen prompting users to enter their username and password. New users have the option to register, which navigates them to a screen where they can register with a username and password or use either Google or Facebook for social login. After proceeding from the login or registration screen, users are presented with a screen asking them to select a secondary authentication method. They can choose from four options: Fingerprint, Email, Phone, and FaceID. This provides two One Time Password (OTP) options (Email and Phone/SMS) and two biometric options (Fingerprint and FaceID). If users select the OTP options, they see a static screen informing them that an OTP has been sent to their email or phone, and they are asked to enter it and press the Submit button. For the prototype, no actual OTP is sent. If users choose a biometric option, they are presented with a screen requesting permission to use Fingerprint or Camera. The permission popup offers two options: Allow and Cancel. Choosing Cancel navigates them back to the screen with the four authentication options. Clicking Accept takes them to a page where they use their Fingerprint or FaceID to log in. For the prototype, these functionalities are represented by a dummy camera shutter button and an on-display fingerprint button that users click. No actual fingerprints or faces are recorded or used. All authentication flow pages lead to the same screen saying ``Thank you for logging in'' to indicate successful login. This page has a Continue button that takes users to their Home Screen.

\subsubsection{Home Screen and Transaction Pages}
The Home Screen presents users with five major elements. At the top, it shows the user's account balance, and below that, the most recent transaction. The top right corner has a profile picture button leading to the Settings page. Below the latest transaction, there are two buttons: one for adding friends with QR codes and another for making a transaction. The transaction button takes users to the Transaction page, where they can enter the amount, select a beneficiary, and send the transaction. For the prototype, the fields to enter the amount and select beneficiaries are static. The Transaction page also shows the user's account balance and has the profile picture icon in the top right corner linking to the Settings page. Clicking the Send button navigates users back to the Home Screen.

\subsubsection{User Profile and Privacy Policies}
The Settings page has four main elements. At the center is a QR code that can be shared with other users to quickly add the primary user as a beneficiary. Below the QR code are three buttons: one for accessing the Social Options feature, one linking to the Privacy Policy, and one for Logout, which navigates the user back to the Login screen. The Privacy Policy button opens a page with a single page of text. The privacy policy template was adapted from PayPal, with references changed to Payment App. This page does not scroll, so the privacy policy text is limited. An `X' button in the top right corner closes the privacy policy page and returns the user to the Settings page. Similarly, the Settings page has an `X' button that closes the page and returns the user to the Home Screen.

\subsubsection{Social Options and Adding with QR Code}
The Social Options button on the Settings page takes users to a page asking for permission to access contacts. The caption at the top reads ``Share Transactions with Friends (off by default).'' For the prototype, clicking Allow or Cancel navigates users back to the Settings page, and no contact information is used. Users can also navigate out of this page by pressing the back button in the top left corner, which returns them to the Settings page. The Add Friends with QR Code button on the Home Screen navigates users to a permissions page requesting camera access. If users deny camera permission, they are directed back to the Home Screen. Clicking Allow takes users to a page where they can scan a friend's QR code. Pressing either the dummy camera shutter button or the `X' button in the top right corner closes the page and navigates users back to the Home Screen.

\begin{figure*}[ht]
     \centering
     \begin{subfigure}[b]{0.2\textwidth}
         \centering
         \includegraphics[width=\textwidth]{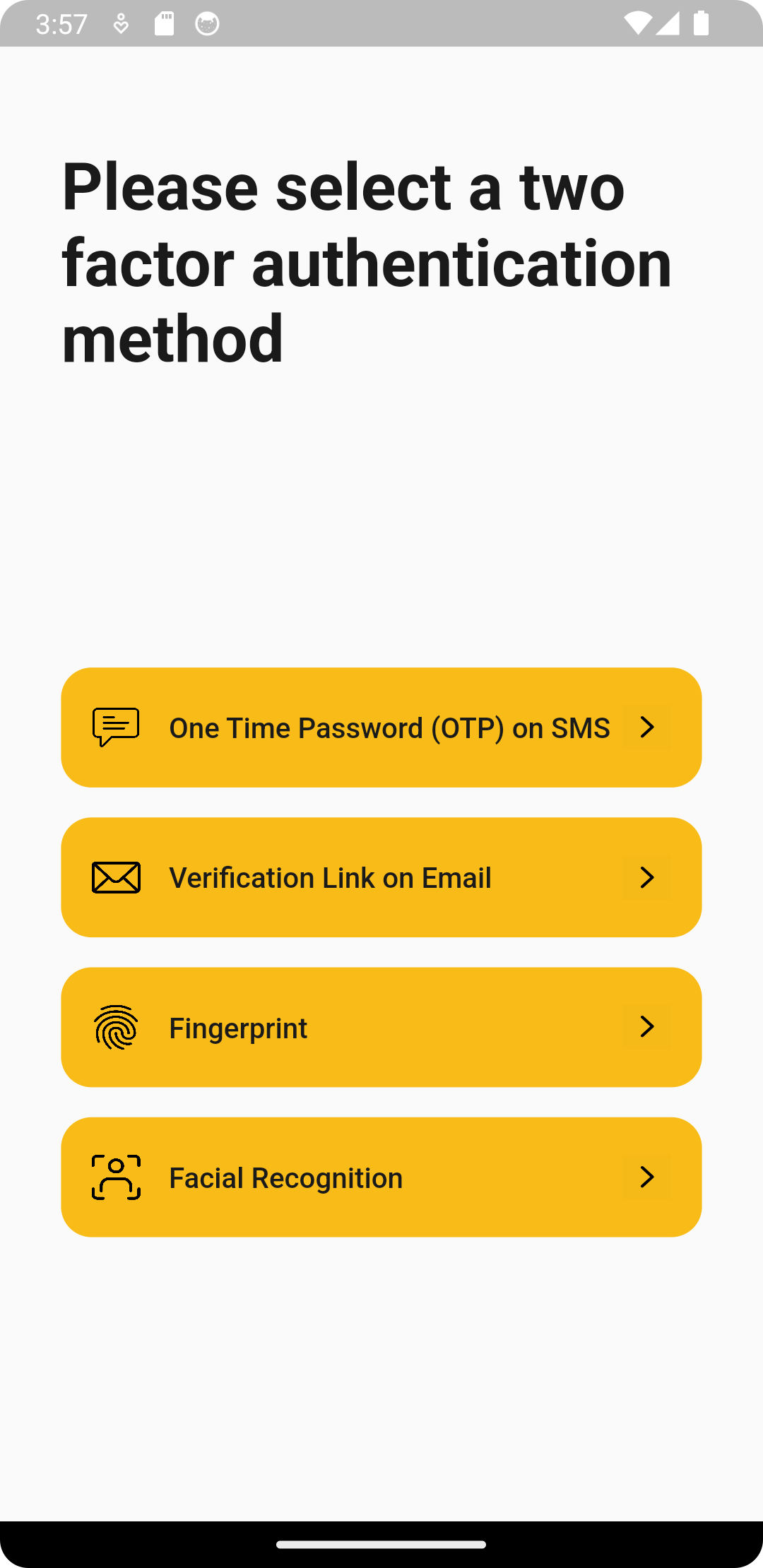}
         \caption{MFA}
         \label{fig:homepage}
     \end{subfigure}
     \hfill
     \begin{subfigure}[b]{0.2\textwidth}
         \centering
         \includegraphics[width=\textwidth]{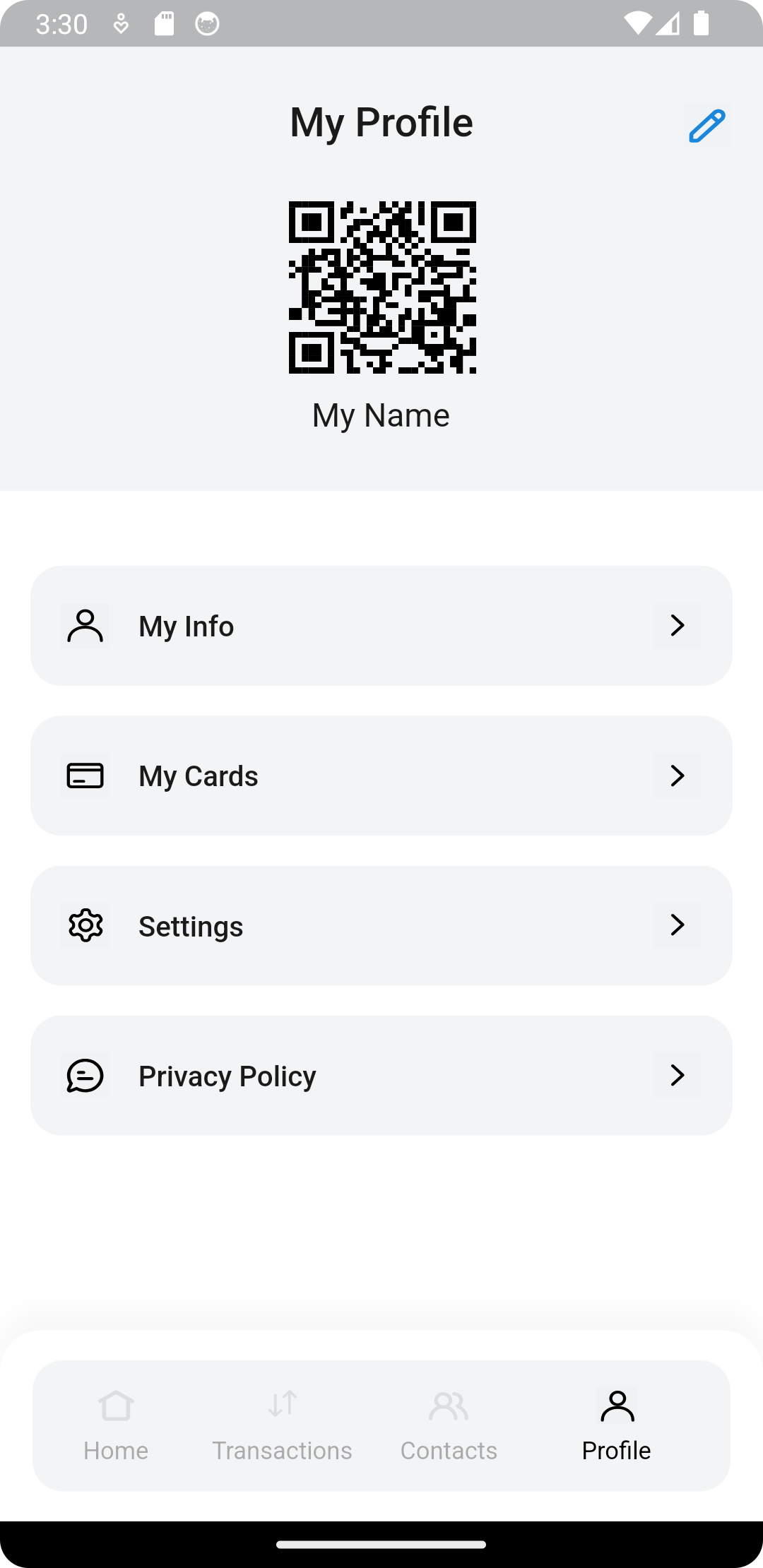}
         \caption{Profile}
         \label{fig:profile}
     \end{subfigure}
          \hfill
    \begin{subfigure}[b]{0.2\textwidth}
         \centering
         \includegraphics[width=\textwidth]{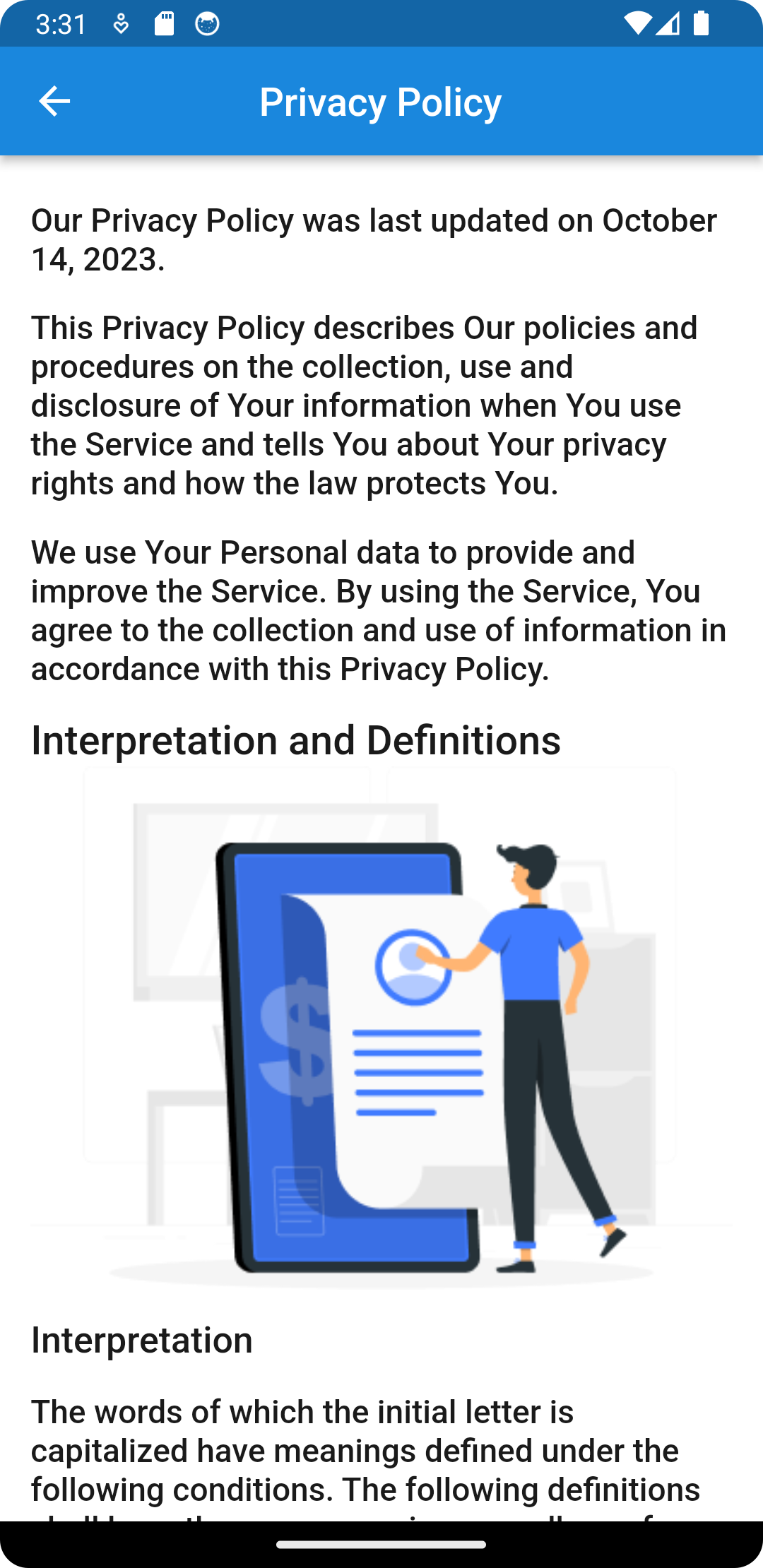}
         \caption{Privacy Policy}
         \label{fig:privacy}
     \end{subfigure}
     \hfill
     \begin{subfigure}[b]{0.2\textwidth}
         \centering
         \includegraphics[width=\textwidth]{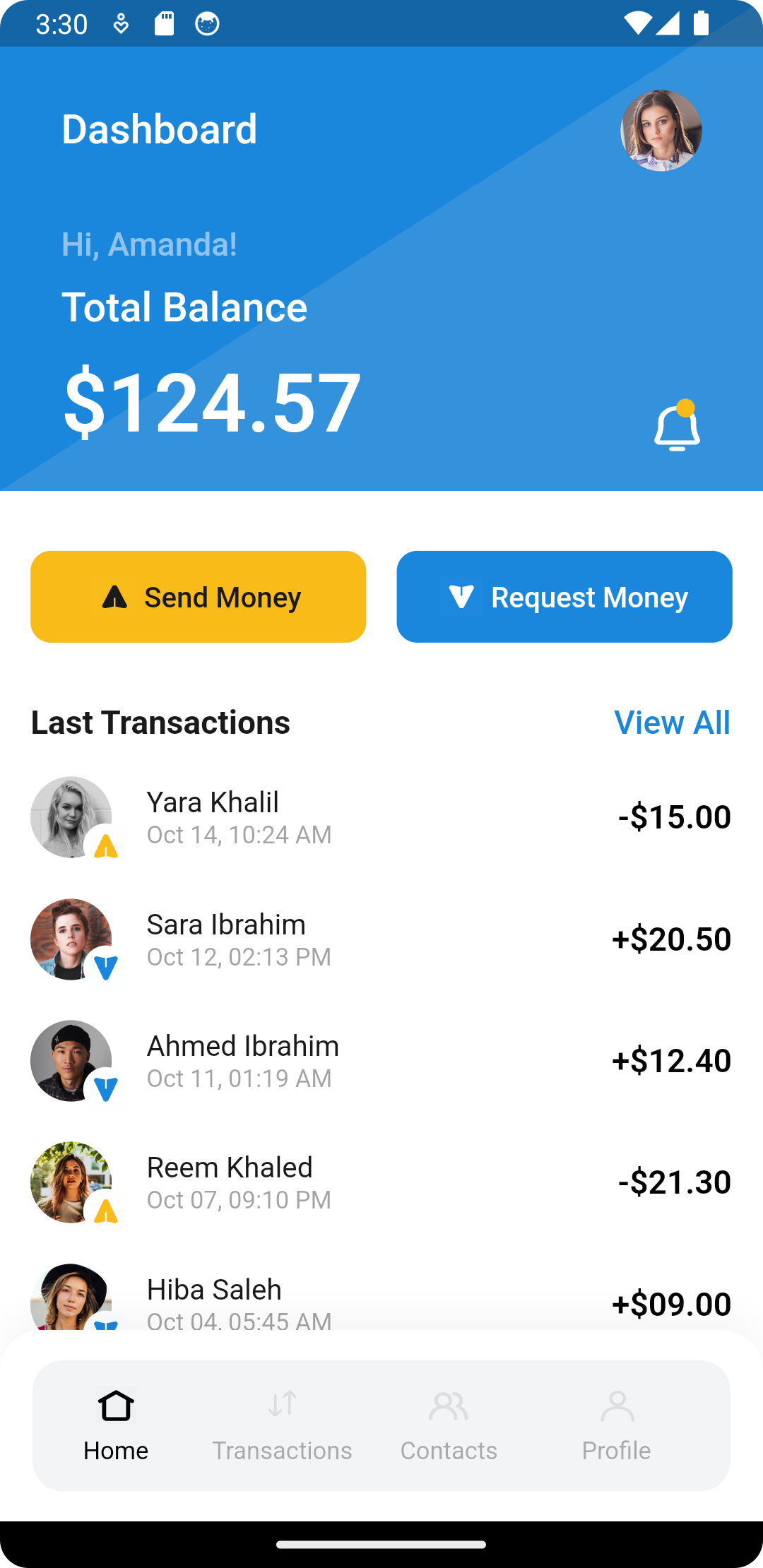}
         \caption{Home Page}
         \label{fig:fingerprint}
     \end{subfigure}
    \caption{Screenshots of Final App}
    \label{fig:app_screenshots}
\end{figure*}

\subsection{User Study}

\subsubsection{Survey Objectives and Development}
The objective of our survey was to understand older adults' perception and assessment of our prototype. This was conducted by showing participants five screenshots from the prototype, which were created based on feedback from our co-design study. Each screenshot was accompanied by questions specifically related to the screenshot or its functionality. Additionally, for each screenshot, we included open-ended questions allowing participants to provide further comments. Lastly, we added a sixth open-ended question at the end of the survey. The survey had a median completion time of $11$ minutes and $19$ seconds.

\subsubsection{Sampling Procedure and Data Collection}
The target population for our user study was older adults aged 60 years and above, following the criteria used in previous studies on older adults~\cite{cham2022too, noah2023safe, zhang2021factors, tang2022never, jin2022used, noah2024evaluating, saka2023safeguarding}. The survey was developed on Qualtrics and deployed on the Prolific crowdsourcing platform, allowing participants to respond remotely. Prolific connects researchers with survey participants for academic and market research studies and has been utilized in numerous prior works~\cite{tazi2023cybersecurity, volkmar2022effects, hertzum2023frustration, zhu2022bias}. To ensure the reliability of our data collection, a power analysis was conducted using z-scores for a $95\%$ confidence interval, a $5\%$ margin of error, and a population size of $79,396,168$ for people aged 60 years and over as reported by the 2022 US Census survey\cite{uscensus}. The calculated power was $365$, and we collected responses from $400$ participants, exceeding the required sample size.

To be eligible for our study, participants had to: (i) be 60 years of age or older, (ii) reside in the United States, and (iii) have experience using at least one e-payment or banking application. We specified the country and age criteria within Prolific to facilitate respondent screening. Additionally, we aimed for a balanced sample with an equal distribution of male and female participants. The criterion of having used an e-payment application was verified within the survey. Participants who successfully completed the survey were compensated with \textdollar$2$ for their time.
\subsubsection{Data Analysis}
To analyze our survey data, we initially collected responses from Qualtrics and the survey data from Prolific. In our Qualtrics form, participants were asked to type their Prolific ID. We used this information to extract Qualtrics responses matching the approved participants list in Prolific. For six approved participants, we manually corrected their participant IDs in the Qualtrics data due to mistyped IDs or IDs typed in an email format. Following this, we conducted analysis of the screenshot-based and demographic questions using Python. 
\subsubsection{Demographics}
Among the $400$ participants, the majority fall within the age range of $60-64$ years, constituting $46\%$ of the participants, followed by those aged $65-69$ years at $30.25\%$. In terms of gender, there is nearly an equal split between male and female participants, with $50\%$ and $49.25\%$, respectively, while a small percentage identifies as non-binary/third gender or prefers not to specify. Regarding education, a significant portion holds a Bachelor's degree ($36.25\%$) or a graduate or professional degree ($27\%$), with smaller proportions having some college experience or technical degrees, high school diplomas, or GEDs. In terms of ethnicity, the majority identify as White or Caucasian ($88.75\%$), with smaller representations from Black or African American, Asian, and other ethnic backgrounds, while some participants prefer not to specify.

\subsection{Mobile Application Development}
We created the $\beta$ version of our mobile application by integrating the feedback from our crowdsourced study into the initial prototype, resulting in a final Android APK. The app was built using Flutter, an open-source framework developed by Google that enables the creation of cross-platform applications using a single codebase. Development took place in Android Studio version Giraffe with Android API level $34$. Flutter utilizes the Dart programming language, with version $3.13.6$-stable employed for this project. Initially, we utilized the code base from 0xayman's GitHub repository, who in turn utilized Hisham Zayadnh's Figma design for the pages. This initial code consisted of static screens, which we modified according to our requirements.

Similar to the prototype, the mobile app comprises multiple interconnected pages incorporating security, privacy, and usability concepts. The app does not connect to any database and utilizes only dummy data to display information on screens. Many screens and flows from the prototype were retained in the mobile app, with additional functionalities introduced. A new landing page displaying the app logo was included to offer users the choice of logging in or registering. The bottom navigation bar was integrated into the app following the login flow, as suggested by participant P8. This navigation bar provides links to: (i) a homepage showing the current balance, recent transactions, and quick buttons for sending or receiving money, (ii) a transaction page listing all transactions, (iii) a contact page with quick actions for sending or receiving money to/from each contact, (iv) send and receive pages displaying appropriate warnings to the user, and (v) other pages for viewing and updating user data such as settings, personal information, and payment cards. The privacy policy page presents concise policies accompanied by graphical elements. Screenshots of the app are provided in Figure~\ref{fig:app_screenshots}.

\section{Results}
\label{ph2_result}
\textbf{Authentication: }
When it comes to signing up for e-payment applications, $91.25\%$ of participants preferred the traditional method of signing up with an email and password. Meanwhile, $11.5\%$ opt for the convenience of signing up through social media platforms like Google or Facebook. Again, participants were allowed to select multiple choices for this question. For preferred choices for secondary authentication after the use of a username and password, $52.25\%$ favor the verification link or One Time Password (OTP) sent via SMS, $38.25\%$ prefer the verification link or OTP sent via email, $29.25\%$ preferred fingerprint authentication, and $27\%$ choose Face ID or Facial Recognition. We note that the majority of participants preferred non-biometric methods, compared to biometric forms of authentication. A small percentage of participants ($0.025\%$) specified that they prefer other methods such as authenticator applications like Google Authenticator or security questions.

\textbf{Privacy \& Permissions: }
Among the participants, the comfort levels regarding permissions granted to e-payment applications vary. A significant $45.25\%$ of users express comfort in allowing notifications, closely followed by location access, with $45\%$ indicating their willingness. A slightly lower proportion, $39\%$, are comfortable granting phone access to these apps, while camera access garners approval from $31.5\%$ of respondents. In contrast, only $11.5\%$ are comfortable with providing access to their contacts, and $8.5\%$ specify that they do not prefer giving any permissions at all. Only $4.75\%$ of users are comfortable granting access to photos and videos.
When it comes to selecting camera permissions, the majority, comprising $43\%$, choose the option of granting access ``While using the app." A significant $36.5\%$ opt for a more temporary access, selecting "Only this time." On the other hand, a notable $20.5\%$ decide to deny camera access entirely, reflecting a substantial portion of users who are cautious about sharing this particular permission.
For location permissions, the majority, constituting $47.75\%$, opt for ``While using the app," indicating a preference for granting location access during app usage but not persistently. A significant $32\%$ choose the more temporary ``Only this time" option, while $17.25\%$ decide to deny location access entirely, showing a notable proportion of cautious users. Only a minor $3\%$ are comfortable with allowing access to their location all the time.

\textbf{Application Design: }
The majority of participants expressed appreciation for the font utilized in the prototype, noting its readability. Additionally, they found the font sizes to be comfortable for reading. Moreover, a smaller segment of participants favored the application logo. Comments also highlighted positive opinions regarding the app's color scheme, with users noting the interface's cleanliness and simplicity. However, some participants mentioned difficulty in reading the font beneath the QR code.

\section{Discussion and Implications}

\subsection{Security Recommendations}
For Two-Factor Authentication (2FA), we suggest employing methods such as OTP via text messages or email links, which are widely favored by older users due to their ease of use and familiarity~\cite{das2018qualitative,das2020smart,patrick2022understanding,reynolds2020empirical}. Strengthening security further, it is essential to implement transaction warnings; clear and prominent alerts before money transfers can effectively deter scams and fraudulent activities. When managing recipients, consider integrating a QR code option for adding contacts. This feature reduces the likelihood of input errors, a common concern among older adults, and enhances overall user satisfaction. Additionally, providing the option for independent account creation within the app empowers older users who may prefer not to rely on social logins for accessibility reasons. Lastly, the integration of standard security icons such as padlocks and shields throughout the app serves to reassure older users about the platform's safety measures. 

\subsection{Privacy Recommendations}When requesting permissions, especially for sensitive data like contacts or photos/videos, it is crucial to prioritize user trust. Clearly explaining the necessity of each permission and providing opt-in choices rather than imposing mandatory permissions is essential~\cite{adhikari2023evolution,adhikari2022privacy}. This approach not only fosters trust but also respects the preferences of users across all age groups. For camera access, offer users control options, with the default setting set to~\lq\lq While using the app.\rq\rq~This aligns with older users' preferences and ensures permissions are app-specific and not persistent. Similarly, with location access, emphasize context-specific usage and respect older users' preferences for limited access. Ensure explanations of the app's location data usage are clear and justified. In terms of privacy policies, it is essential to craft concise and clear documents. Consider breaking them into easily digestible sections or bullet points for improved comprehension, particularly among older users. Enhance the readability of privacy policies with graphical elements such as icons, images, or infographics. These visual aids make the policies more engaging and understandable for older adults. 

\subsection{Usability Recommendations}
In addition to our platform's security and privacy guidelines, we have tailored further recommendations to enhance usability and accessibility specifically for older adults. These recommendations focus on design elements and language choices that accommodate the diverse needs of this demographic. Firstly, we suggest incorporating larger font sizes and high-contrast text to improve readability, considering potential visual impairments associated with aging. Additionally, utilizing a color palette with sufficient contrast and avoiding sole reliance on color for conveying information can aid users with visual impairments in navigating the platform effectively. Maintaining a simple and intuitive design is essential, minimizing clutter and complexity to cater to older adults who may prefer straightforward interfaces. Clear and well-organized navigation menus should be implemented to help them easily locate desired features within the app. Furthermore, communication should employ clear and concise language, avoiding technical jargon that might be unfamiliar. For instance, using terms like ``recipient" instead of ``beneficiary" and providing explanations for acronyms like "SMS" (text message) can enhance comprehension. 

\section{Acknowledgement}
We would also like to thank our participants for this study and acknowledge the Inclusive Security and Privacy-focused Innovative Research in Information Technology (InSPIRIT) Lab for supporting this work. Any opinions, findings, conclusions, or recommendations expressed in this material are solely those of the author.

\bibliographystyle{plain}
\bibliography{Buildsec}

\end{document}